\documentclass[12pt]{article}
\usepackage{latexsym}
\usepackage{axodraw}
\usepackage{epsfig}
\begin{document}
\begin{center} {\Large \bf  Ultraviolet Fixed Points in Gauge and SUSY\\[0.5cm]
 Field Theories in Extra Dimensions
} \vspace{1cm}

{\large \bf D.I.Kazakov} \vspace{0.7cm}

{\it Bogoliubov Laboratory of Theoretical Physics, Joint
Institute for Nuclear Research, Dubna, Russia \\[0.2cm] and\\[0.2cm]
Institute for Theoretical and Experimental Physics, Moscow,
Russia}
\end{center}

\begin{abstract}
We consider gauge field theories in $D>4$ following the Wilson RG
approach and show that they possess the ultraviolet fixed points
where the gauge coupling is dimensionless in any space-time
dimension.  At the fixed point the anomalous dimensions of the
field and vertex operators are known exactly. These fixed points
are nonperturbative and correspond to conformal invariant
theories. The same phenomenon also happens in supersymmetric
theories with the Yukawa type interactions.
\end{abstract}

\section{Introduction}

Nowadays it became popular to consider theories in extra
dimensions as possible candidates for models of physics beyond the
Standard Model. (See e.g. Ref.\cite{Alt,BHN} and references
therein.) One may wonder whether this extra dimensional theory can
be considered as a consistent QFT in any sense. Since by general
power counting it is nonrenormalizable, it looks hardly possible.

One way to consider an extra dimensional theory is the
Kaluza-Klein approach. In this case, one takes the Fourier
transform over the extra dimensions and obtains an infinite tower
of states with quantized masses. Then one has to sum over all the
states. This sum is usually divergent and a special prescription
is needed to regularize it. Following this approach the
divergences in D=5 SUSY theory have been studied
in~\cite{Ant,GN,KK}  for the scalar effective potential. Some
cancellations of UV divergences have been found. Doubtfully,
however, that this approach solves the problem of
nonrenormalizability in extra dimensions.

In principle, there is a chance that all the UV divergences cancel
each other, like it takes place in N=4,\ 2 and even N=1 SUSY field
theories in D=4~\cite{finite}, and one has a consistent theory.
This possibility has been studied in the
literature~\cite{MS,T,HS,K}. Though at lower orders the
divergences indeed cancel on shell~\cite{MS,T,K},  in higher
orders they may well appear being unprotected by any
symmetry~\cite{HS}.

In what follows we first remind the situation with the UV
divergences in SUSY gauge theories in extra dimensions in the
lowest order and then discuss an alternative approach based on the
Wilson renormalization group fixed points. The latter one is
applied to the usual as well as supersymmetric gauge theories and
exploits the nonperturbative RG fixed points for $D>4$.

\section{One-loop UV divergences in SUSY theories \protect\\ for arbitrary $\mathbf D$.}

Consider the one-loop vacuum polarization diagram in a non-Abelian
gauge theory. It can be evaluated in arbitrary dimension using the
technique of dimensional regularization. The result in the
background field formalism is (we omit the transverse polarization
tensor)
\begin{eqnarray}
  \Pi(p^2)& = &  (-)^{[D/2]}\frac{\Gamma(2-D/2)\Gamma^2(D/2)}{\Gamma(D)} \times
  \nonumber\\
    \left\{ \right.&-&
   \left[2\frac{8(D-1)-D'}{D-2}+\frac{(D-4)(D-1)\alpha(8-\alpha)}{2(D-2)}+\frac{4}{D-2}\right]
   C_2(G) \nonumber\\
   &+&\left.2^{[D'/2]}T(R)+\frac{4}{D-2}T(R)\right\}\frac{1}{(p^2)^{2-D/2}}, \label{vac}
\end{eqnarray}
where $D$ is the dimension of integration and $D'$ is the
dimension of the fields corresponding  to the Lorentz algebra. We
present the result in an arbitrary $\alpha$-gauge ($\alpha=0$
corresponds to the Feynman gauge). The square bracket contains the
gauge and ghost field contribution,  and then follows those of
spinor and scalar fields.

Taking $D=4-2\varepsilon$ in eq.(\ref{vac}) one can reproduce the
result for the logarithmic, quartic and sextic divergences in
$D=4, 6$ and $10$, respectively. The singular part is proportional
to
\begin{equation}\label{s}
  -(26-D')C_2(G)+2^{[D'/2]}T(R)+2T(R).
\end{equation}
This is a gauge invariant expression of invariant operator
$F^2_{\mu\nu}$.

Consider eq.(\ref{s}) in particular cases corresponding to SUSY
gauge theories in various dimensions taking the proper sets of the
matter fields. The results are summarized  below
$$\begin{array}{llll}
D'=4 \ \ \ & N=1\ \ \ & -22C_A+4C_A+4T_R+2T_R & =-6(3C_A-T_R),\\
  &N=2& -22C_A+4C_A+6C_A+12T_R &=-12(C_A-T_R),\\
  & N=4&-12C_A+12C_A&=0,\\ &&&\\
  D'=6&N=1& -20C_A+8C_A+8T_R+4T_R&=-12(C_A-T_R),\\
  &N=2& -12C_A+12C_A&=0,\\ &&&\\
  D'=10&N=1&-16C_A+16C_A&=0.
\end{array}$$
One can see that when the matter field representations are chosen
in a proper way, the leading divergences indeed cancel each other.
Note that the $N=1\ D=10$ case coincides with the $N=2\ D=6$ and
$N=4\ D=4$ ones and the $N=1\ D=6$ case coincides with the $N=2\
D=4$ one as expected.

Return to logarithmic divergences in higher dimensions. Take $D=6$
for definiteness. Due to the background field gauge invariance the
divergent structures in the one-loop order can take one of the
following forms:
\begin{eqnarray}
I_1&=&Tr D_\rho F_{\mu\nu} D_\rho F_{\mu\nu}\ , \label{str} \\
I_2 &=&Tr D_\mu F_{\mu\nu} D_\rho F_{\rho\nu}\ , \nonumber \\
I_3&=&  Tr D_\rho F_{\mu\nu} D_\mu F_{\rho\nu}\ , \nonumber \\
I_4&=& Tr F_{\mu\nu} F_{\nu\rho} F_{\rho\mu}\ . \nonumber
\end{eqnarray}
However, these invariants are not independent. Due to the relation
$[D_\mu,D_\nu]=F_{\mu\nu}$ and the Bianchy identity $D_\mu
F_{\nu\rho}+D_\rho F_{\mu\nu}+D_\nu F_{\rho\mu}=0$, one has only 2
independent structures and  can choose any of them. We take the
first two. Then calculating the diagrams and extracting the
contribution to two independent Lorentz structures one can find
the coefficients  of them. The result is
\begin{equation}
\frac{T_R-C_A}{3}\ Tr D_\mu F_{\mu\nu} D_\rho
F_{\rho\nu}.\label{gauge}
\end{equation}
One finds that the result for ALL the structures is proportional
to $\sum T(R) - C_2(G)$, like  for the quadratic divergences, and
vanishes {\it off-shell}. Due to the fact that all the structures
vanish we claim that all the one loop divergences in the gauge
sector cancel for $\sum T(R) = C_2(G)$!

However, unlike the quadratic divergences, this result is
gauge-dependent. In an arbitrary $\alpha$-gauge eq.(\ref{gauge})
looks like
\begin{equation}
\frac{T_R-C_A(1+\alpha-\alpha^2/8)}{3}\ Tr D_\mu F_{\mu\nu} D_\rho
F_{\rho\nu}.\label{gauge2}
\end{equation}
and the cancellation is not  obvious anymore.

To get a gauge invariant statement, one has to go on-shell, i.e.
to use the equations of motion. For the pure gauge case they are
\begin{equation}\label{eq}
  D_\mu F_{\mu\nu}=\bar\lambda \gamma^\nu \lambda, \ \ \ \ \hat D
  \lambda=0,
\end{equation}
where $\lambda$ is the gaugino field. Collecting the terms of
effective action which transform into one another due to the
equations of motion one has
$$...(D_\mu F_{\mu\nu})^2 + ...\bar\lambda \gamma^\nu D_\mu F_{\mu\nu}\lambda +
... (\bar\lambda \gamma^\nu \lambda)^2 = 0\ \ \ !\ ,$$ where the
dots stand for the know coefficients. That is one finds
cancellation of the logarithmic divergences {\it on-shell} in any
gauge.

In higher loops the following statements are valid:
\begin{enumerate}
\item The on-shell finiteness of the $D=6\ N=1$ SUSY gauge theory
is true in two loops as well. This has been checked by explicit
calculation in components~\cite{MS,T};
\item Within the (constrained) superfield formalism it is possible
to show that the allowed invariants vanish on-shell up to 2 loops.
However, in higher loops the nonvanishing invariants
exist~\cite{HS}. The coefficients are not calculated but there is
no known symmetry that might protect them.
\end{enumerate}
Thus, our main conclusion is not optimistic: there is no big
chance for the cancellation of logarithmic divergences for $D>4$
even on-shell, i.e. the theory remains {\it perturbatively
nonrenormalizable}.

\section{Nonperturbative fixed point in gauge theories \protect\\
for $\mathbf{D>4}$.}

We turn now to an alternative idea and look for nonperturbative
possibilities to construct a viable higher dimensional theory. We
follow the so-called Wilson Renormalization Group
approach~\cite{Wilson}, only not in a scalar theory but in a gauge
one. Our treatment of nonrenormalizable interactions follows that
of M.Strassler~\cite{S,S2}.

Consider first the usual gauge theory in $D$ dimensions
\begin{equation}\label{l}
  {\cal L}=-\frac 14 Tr F^2_{\mu\nu}, \ \ \
  F_{\mu\nu}=\partial_\mu A_\nu - \partial_\nu A_\mu + g
  [A_\mu,A_\nu].
\end{equation}
The fields and the coupling have the following canonical
dimensions:
$$[A]=\frac{D-2}{2}\ ,\ \  \ \ [F]=\frac D2 \ ,\ \  \ \ [g]=2-\frac D2. $$
This means that $D=4$ is the critical dimension for the gauge
interaction: the coupling here is dimensionless, the operators are
marginal and the theory is renormalizable in a usual sense.

A dimensional analysis implies consideration of the  dimensionless
quantity
$$\tilde g \equiv g \mu^{D/2-2} \ \ \ \ \Rightarrow \ \ \ \ [\tilde g] = 0,$$
where $\mu$ is some scale.\footnote{Remind in dimensional
regularization~\cite{H} \ $g_{Bare}=g\mu^\varepsilon$ in
$D=4-2\varepsilon$.} Now one can go  to the critical dimension
$D=4$ where the theory is renormalizable, and write down the RG
equation for $g$
\begin{equation}
  \mu\frac{d}{d\mu}g =   g(\frac 12 \gamma_A),
\end{equation}
where $\gamma_A$ is the gauge field anomalous dimension in the
background field gauge. This gives, following Wilson's approach,
the RG equation for $\tilde g$ which we consider in an arbitrary
dimension $D$ via the analytical continuation
\begin{equation}\label{nu}
 \mu\frac{d}{d\mu}\tilde g =   \tilde g\frac 12 \gamma_A+\tilde g(\frac
  D2-2)=\frac{\tilde g}{2}(\gamma_A+D-4).
\end{equation}
Eq.(\ref{nu}) has a fixed point. In fact, two of them
$$\begin{array}{l}
1) \ \ \tilde g=0 \ \ \rightarrow g=0, \ \ \gamma_A=0, \\
2) \ \ g=g^*, \ \ \gamma_A=4-D.
\end{array}$$
The first one is trivial, this is the so-called Gaussian fixed
point. It is perturbative. The second one is nonperturbative, it
is the so-called Wilson-Fisher fixed point\cite{Wilson}. The
anomalous dimension here is not small, it is integer. It is
achieved at the value of the coupling which is unknown, though the
value of the anomalous dimension is known {\it exactly}. Since the
anomalous dimension in gauge theories, contrary to the scalar
case, is negative, the fixed point of the second kind exists for
$D>4$. Remind that in scalar theories it exists for $D<4$: one
takes $D=4-\epsilon$, where $\epsilon \to 1$ or $2$ and performs
the so-called $\epsilon-$expansion~\cite{Wilson}. In the case of a
scalar theory the FP is IR stable, while in a gauge theory it is
UV stable (see Fig.1).
\begin{figure}[ht]
\begin{center}
\begin{picture}(200,130)(30,-30)\Text(-5,80)[]{$\beta(g)$}\Text(100,0)[]{$g$}
\Text(0,0)[]{$0$}\Text(50,0)[]{$g^*$}\Vertex(10,10){2}\Vertex(43,10){2}
\Line(0,10)(90,10)\Line(10,0)(10,90) \Curve{(10,10)(20,-0)(65,60)}
\Line(150,10)(230,10)\Line(160,0)(160,90)\Text(140,80)[]{$\beta(g)$}\Text(240,0)[]{$g$}
\Curve{(160,10)(170,20)(215,-5)}
\Text(150,0)[]{$0$}\Text(200,0)[]{$g^*$}\Vertex(160,10){2}\Vertex(205,10){2}
\Text(40,-30)[]{Scalar Theory\ ($D<4$)}\Text(200,-30)[]{Gauge
Theory\ ($D>4$)} \Text(75,30)[]{IR FP}\Text(215,30)[]{UV FP}
\ArrowLine(65,22)(50,15)\ArrowLine(215,22)(205,15)
\end{picture}\caption{}
\end{center}
\end{figure}

Such a fixed point in a gauge theory within the
$\epsilon$-expansion has been  advocated in ref.\cite{P}. Some
additional supporting arguments in favour of the fixed point in 5
dimensions come from the lattice calculations~\cite{Lat}. At last,
there is also very useful analogy between the gauge theory and the
nonlinear sigma-model. The latter has a critical dimension equal
to 2 and is asymptotically free there. One can go above the
critical dimension with the help of the $2+\epsilon$ expansion.
Then the theory has a fixed point in the leading order~\cite{PS}
which is also true within the 1/N expansion performed directly in
three dimensions~\cite{A}.

 Consider the properties of the fixed point $\# 2$. Let us
calculate the dimensions. One has for the field
$$[A]=\frac{D-2}{2}+\frac
12\gamma_A=\frac{D-2}{2}+\frac{4-D}{2}=1$$ in any $D$. To
calculate the dimension of the coupling, one has to consider the
vertex $g\partial A[A,A]$ which gives
$$D=[g]+1+3[A]+\gamma_V.$$\
Since $\gamma_V=-\gamma_A$ in the background gauge, one obtains
$$[g^*]=D-4-\gamma_V=D-4+\gamma_A=0  \ \ \ \ \mbox{in any D}\ \ \mathbf{!}$$

Thus, one has a dimensionless coupling at the fixed point that
means renormalizability. {\it The theory at the fixed point is
perturbatively nonrenormalizable, but nonperturbatively
renormalizable!} (cf Ref.\cite{S2}). The existence of a
renormalizable field theory beyond PT relies, in the sense of
statistical physics, on the existence of a fixed point~\cite{ZJ}.

How  can one understand this statement in terms of Feynman
diagrams? Compare the two fixed points, the Gaussian one and the
nonperturbative one

\begin{center}
\begin{tabular}{l|l}
$g=0$ & \ \ \ \ $g=g^*$ \\ & \\$\widehat{AA} \sim
\frac{\displaystyle 1}{\displaystyle (x^2)^\frac{D-2}{2}}$\ \ \ \
\ \ \  &\ \ \ \ $\widehat{AA} \sim \frac{\displaystyle
1}{\displaystyle (x^2)^1}$ \\ & \\$\int\frac{\displaystyle d^Dx
e^{ipx}}{\displaystyle (x^2)^\frac{D-2}{2}}\sim
\frac{\displaystyle 1}{\displaystyle p^2}$&\ \ \ \
$\int\frac{\displaystyle d^Dx e^{ipx}}{\displaystyle (x^2)}\sim
\frac{\displaystyle 1}{\displaystyle (p^2)^\frac{D-2}{2}}$
\end{tabular}
\end{center}
Thus, for instance, for $D=6$ at the non-Gaussian fixed point the
propagator behaves like $1/p^4$, i.e. much faster than in the
usual case.

One can consider the diagrams with modified Feynman rules taking
into account the anomalous dimensions. This corresponds to
infinite summation of subgraphs. For the gauge propagator one has
by power counting
\begin{center}
\begin{picture}(200,130)(50,-30)
\Gluon(10,70)(31,70){2}{3}\Line(28,77)(37,70)\Line(70,77)(63,70)
\GlueArc(50,70)(16,360,0){2}{12}\Gluon(68,70)(90,70){2}{3}
\Text(195,70)[]{$\frac{\displaystyle
D+2-2+2\gamma_V}{\displaystyle 4-2\gamma_A} \ \Rightarrow \ D-4,$}
\Gluon(10,20)(31,20){2}{3}\Gluon(50,5)(50,35){2}{3}\Line(28,27)(37,20)
\Line(70,27)(63,20)\Line(45,30)(55,30)\Line(45,10)(55,10)
\GlueArc(50,20)(16,360,0){2}{12}\Gluon(68,20)(90,20){2}{3}
\Text(230,20)[]{$\frac{\displaystyle
2D+4-2+4\gamma_V}{\displaystyle 10-5\gamma_A} \ \Rightarrow \
2D-8+\gamma_A=D-4,$} \Text(40,-20)[]{.\ .\ .\ .\ .\ .\ .}
\Text(245,-20)[]{$\Rightarrow \ \ \ D-4$.}
\end{picture}
\end{center}
Hence, one has the same power in any loop, that is
renormalizability. This is the consequence of dimensionless
coupling at the fixed point.

One can try to construct an effective Lagrangian that describes
these diagrams. In $D=6$, as it is suggested by the one-loop
calculation (\ref{gauge}) and the behaviour of the propagator, it
may be
\begin{equation}\label{eff}
  {\cal L}_{eff}\sim \ Tr (D_\mu F_{\mu\nu})^2.
\end{equation}
The effective Lagrangian (\ref{eff}) has some remarkable
properties
\begin{itemize}
\item It has no scale, the coupling is dimensionless;
\item It is scale (conformal) invariant;
\item The exact anomalous dimensions of the field and vertices are
taken into account;
\item It is vanishing on-shell ($D_\mu F_{\mu\nu}=0$).
\end{itemize}
At first sight, the effective Lagrangian (\ref{eff}) contains
higher derivatives, and hence, ghosts. However, it is not clear
for us how to define the spectrum of effective theory: is it the
spectrum of the original Lagrangian or may be some new fields are
adequate in this case?

\section{Nonperturbative fixed point in  SUSY theories \protect\\
for $\mathbf{D>4}$.}

A similar phenomenon takes place in SUSY gauge theories. Again we
start at the critical dimension $D=4$ and use $N=1$ superfields.
Strictly speaking, they are $D=4$ superfields; however, component
notation is more cumbersome and what we really need are the
renormalizations in a critical dimension. So the superfield
formalism here is not rigorous but useful. Remind also that
supersymmetry is possible only at integer dimensions with $D\leq
10$ if one restricts the maximum spin=1.

The SUSY Lagrangian looks like (we omit the gauge fields for the
moment)
\begin{equation}\label{lag}
  {\cal L}=\int d^4\theta \ \bar\Phi_i \Phi_i + \int d^2\theta \ {\cal W} +h.c., \
  \ \ \
  {\cal W}=y \Phi_1\Phi_2\Phi_3.
\end{equation}
Calculating the dimensions of the fields and the Yukawa coupling
$y$, one has
$$[{\cal L}]=D, \ \ [d\theta]=1/2,\ \ [{\cal W}]=D-1,$$
$$[\Phi]=\frac{D-2}{2}, \ \ [y]=D-1-3\frac{D-2}{2}=2-D/2.$$
Now we proceed as above. Introduce a dimensionless quantity
$\tilde y=y\mu^{D/2-2}$ and write the RG equation for $y$ in $D=4$
\begin{equation}\label{y}
  \mu\frac{d}{d\mu}y=y(\frac 12 \gamma_1+\frac 12 \gamma_2+\frac 12
  \gamma_3),
\end{equation}
where $\gamma_i$ is the anomalous dimension of the matter field
$\Phi_i$. We use here the nonrenormalization theorem in $D=4$
which states that the anomalous dimension of the vertex is zero.

This allows us to get the RG equation for $\tilde y$
\begin{equation}\label{ty}
  \mu\frac{d}{d\mu}\tilde y=\tilde y(\frac 12 \gamma_1+\frac 12 \gamma_2+\frac 12
  \gamma_3)+\tilde y(D/2-2)=\frac{\tilde y}{2}(\gamma_1+\gamma_2+\gamma_3+D-4).
\end{equation}

This equation has two fixed points~\cite{S}\footnote{ For a scalar
SUSY theory in $D<4$ this nonperturbative fixed point was earlier
used in Ref.\cite{SARV} to describe the self-avoiding random
walk.}
$$\begin{array}{l}
1) \ \ \tilde y=0 \ \ \rightarrow y=0, \ \ \gamma_i=0, \\
2) \ \ y=y^*, \ \ \gamma_i=(4-D)/3.
\end{array}$$
One can see that for $D>4$ the second fixed point requires the
anomalous dimension to be negative. This is only possible in gauge
theories. Hence, in fact one has to consider the gauge invariant
SUSY theory where the nontrivial fixed point is ($g^*,y^*$). At
this point the dimension of the Yukawa coupling is
$$[y^*]=D-1-3\frac{D-2}{2}-\frac{\gamma_1+\gamma_2+\gamma_3}{2}=0
\ \ \ \ \mbox{in any D}\ \ \mathbf{!}$$

Thus, again, we get a theory that is perturbatively
nonrenormalizable, but nonperturbati\-ve\-ly renormalizable at the
nontrivial fixed point. At this point a theory possess all the
properties mentioned above.

There is, however, one subtlety with supersymmetry. While
supersymmetric gauge theory exists for $D\leq 10$, it is known
that superconformal algebra, which we assume to be realized at the
fixed point, is only possible for $D\leq 6$. This is due to the
classification of all possible superconformal algebras given in
ref.~\cite{Nahm}. At the same time due to the vanishing beta
function the conformal anomaly should also vanish at the fixed
point so the theory is scale invariant. Hence, we find some kind
of discrepancy here\footnote{I am grateful to E.Witten for
pointing out this problem to me.}. There might be two ways out of
it: either the fixed point solution is not valid for some reason
above D=6, or the full superconformal algebra is not realized at
the fixed point. This remains an open question so far.

One may wonder whether this nontrivial fixed point is reachable.
To see this, consider an $N=1$ SUSY gauge theory and take the
all-loop NSVZ $\beta$-function~\cite{NSVZ}. Extracting the
$\gamma_A$ one finds
$$\gamma_A=2\alpha\frac{T_R-3C_A-\frac{2}{r}\sum C_R
\gamma_R}{1-2C_A\alpha},
$$ where $\alpha\equiv g^2/16\pi^2$.

For a pure SUSY Yang-Mills case one has the equation
$$\gamma_A=2\alpha\frac{-3C_A}{1-2C_A\alpha}= 4-D.$$
The solution is
$$\alpha^*=\frac{D-4}{D-1}\ \frac{1}{2C_A}.$$
This value is smaller than the pole value $\alpha_{pole}=1/2C_A$.
In particular, in $D=6$ one has $\alpha^*=1/5C_A$, as shown in
Fig.2.
\begin{figure}[ht]
\begin{center}
\begin{picture}(100,130)(30,-10)
\Line(0,70)(120,70)\Line(0,0)(0,110) \DashLine(70,0)(70,100){2}
\DashLine(0,40)(40,40){2}\DashLine(40,40)(40,70){2}
 \Curve{(0,70)(40,40)(58,20)(67,0)}
\ArrowLine(85,80)(71,71)\ArrowLine(25,80)(40,71)\Text(135,70)[]{$\alpha$}
\Text(-10,100)[]{$\gamma_A$}\Text(-10,70)[]{0} \Text(-10,40)[]{-2}
\Text(20,92)[]{$\frac{1}{5C_A}$}\Text(90,92)[]{$\frac{1}{2C_A}$}
\end{picture}\caption{}
\end{center}
\end{figure}
Thus, the fixed point seems to be quite reachable.

\section{Conclusion}

Summarizing the analysis of the gauge and SUSY field theories in
higher dimensions from the point of view of their
renormalizability and consistency, we come to the following
conclusions
\begin{enumerate}
  \item[-] Perturbative finiteness in $D>4$ seems not to be valid;
  \item[-] Within the Wilson RG approach the nontrivial
  nonperturbative fixed points may lead to nonperturbative
  renormalizability;
  \item[-] These theories may be related to PT
  renormalizable effective models which have to be found;
  \item[-] At the fixed point the theory possesses the conformal
  invariance, and the anomalous dimensions are known exactly.
\end{enumerate}

These observed fixed points may be related to those originated
from the  string dynamics  known already for several years for D=5
and 6 (see ref.\cite{SW} and references therein). We use here the
more familiar language that is close to statistical physics and
critical phenomena. In a sense we give an explicit example of a
local field theory  with non-trivial fixed points thus
strengthening the claim (based on string theory) that exist field
theories that flow to non-trivial fixed points in more than 3
dimensions~\cite{SW}. It would be very useful to our mind to find
explicit link between the two approaches.

\section*{Acknowledgements} I would like to thank M.Strassler,
M.Shifman, V.Miransky, T.Jones, I.Jack, V.Rubakov, and A.Slavnov
for useful discussions. I am grateful to M.Shifman for the
invitation to the Conference "Continuous Advances in QCD-2002"
where this investigation was started. Financial support from RFBR
grants \# 02-02-16889 and \# 00-15-96691 is kindly acknowledged.

\end{document}